\documentclass[preprint,12pt]{elsarticle}

\usepackage{amssymb}

\usepackage{graphicx}

\journal{Optik}

\begin{document}

\begin{frontmatter}



\title{Nonlinearity managed dissipative solitons}


\author{Fatkhulla  Kh. Abdullaev$^{1,2}$, Sadulla Sh. Tadjimuratov$^1$, Abdulaziz A. Abdumalikov$^2$}

\address{
$^1$Physical-Technical Institute of the Uzbekistan Academy of Sciences, Ch. Aytmatov str.,
2-B, 100084, Tashkent-84,  Uzbekistan.\\
$^2$Theoretical Physics Department, Faculty of Physics, National University of Uzbekistan,
University street 4, 100174, Tashkent-174, Uzbekistan
}

\begin{abstract}
   We study the dynamics of localized pulses in the complex cubic-quintic Ginzburg-Landau
(GL) equation with strong nonlinearity management. The generalized complex GL equation,
averaged over rapid modulations of the nonlinearity, is derived. We employ the
variational approach to this averaged equation to derive a dynamical system for the
dissipative soliton parameters. By using numerical simulations of this dynamical system
and the full GP equation, we show that nonlinearity-managed dissipative solitons exist in
the model. The parameter regions for stabilization of exploding soliton are obtained.
\end{abstract}



\begin{keyword}
Complex Ginzburg-Landau equation, dissipative solitons, pulsating solitons,
exploding solitons, nonlinearity management, variational approach.



\end{keyword}

\end{frontmatter}


\section{Introduction}
\label{intro}

   Dissipative solitons attract much attention nowadays due to a general importance for
the theory of nonlinear waves and applications, such as mode-locked lasers, optical
devices with gain/loss parameters, and other areas~\cite{DissSol,Grelu12}. In optics, a
wide range of systems possessing dissipative solitons is described by the complex
Ginzburg-Landau (GL) equation~\cite{DissSol}. As many studies show, this equation has a
rich variety of solutions, including stationary solitons, pulsating solitons, exploding
solitons, and fronts~(see e.g. \cite{AST}).

   Pulsating solitons represent localized solutions with periodically or
quasi-periodically varying amplitudes and widths. Pulsating solitons in the
complex GL equation were found numerically in works~\cite{AST,Deissler,SGGA}. Such
solitons has been observed experimentally in fiber lasers~\cite{SGGA}. It was
shown that pulsating solitons are stable limit cycles (LC) of a finite dimensional
dynamical system~\cite{Tsoy}.

    Exploding solitons(ES) are solitons with abrupt spreading and explosions of the
pulse. Usually explosions are different from each other. After an explosion, a pulse has
almost the same shape, though the pulse position can change. An interval between
explosions and a shift of the pulse position seem to be random. Exploding solitons were
observed experimentally in a sapphire laser~\cite{Cundiff}, in a mode-locked fiber
laser~\cite{Runge}, and in an anomalous dispersion fiber laser~\cite{Wu19}.

   Different schemes for stabilization and control of these pulses have been suggested.
One of them is to take into account the higher-order effects. This leads to a formation
of periodic, non-chaotic one-side explosions~\cite{Gurevich}, and it can provide the
shape stabilization of ESs~\cite{Latas}. Also, the dispersion management (DM) method,
based on the rapid and strong variations of the dispersion on the evolutional variable,
can be used to generate stable pulses with the enhanced energy. For dispersion-managed
optical systems, some numerical results were obtained in works~\cite{SotoOE,DissSol},
while an experiment was performed in Ref.~\cite{OptExpr04}. In works~\cite{AHI,Biondini},
the averaged GL equation has been derived. It was shown that the dispersion-managed GL
equation is well approximated by the averaged over dispersion variations of nonlinear
Schr\"{o}dinger equation with the gain/loss terms of the same form as the
original GL equation. The moments method has been applied to this problem in
Ref.~\cite{Tur}. Mathematical aspects of the theory of dispersion managed optical solitons
are discussed in the work ~\cite{Biswas}.

   Another method that can be applied for soliton stabilization is the nonlinearity
management (NM),which is based on the rapid and strong variations of the nonlinearity.
For conservative systems, it was shown that nonlinearity management has many advantages
for a generation of stable solitons in optical communication systems, in fiber
lasers~\cite{Ilday}, and for generation of stable 2D spatial
solitons~\cite{Berge,Towers}. In attractive Bose-Einstein condensates (BECs), described
by the focusing multi-dimensional nonlinear Schr\"{o}dinger equation, the NM
stabilization of solitons in BECs has been developed in Refs.~\cite{AK2001,ACMK,SU,Mont}.
Therefore, it is of interest to study a possibility of dynamical stabilization of
solitons in the complex cubic-quintic Ginzburg-Landau equation by the nonlinearity
management.

   In this work, we consider the dynamics of dissipative solitons in the complex
cubic-quintic Ginzburg-Landau equation with a strong nonlinearity management. Such
variations can be realized, for example, in a fiber ring laser with two-step (focusing
and defocusing ) variations of nonlinearity.

   The structure of the paper is as follows. In the Section~\ref{sec:aver}, we describe a
nonlinearity management model, and derive the generalized complex GL equation, averaged
over the period of rapid modulations of nonlinearity. This complex GL equation is
analyzed using the variational approach and the system of equations for dissipative
soliton parameters is derived in Sec.~\ref{sec:breath}. Results of numerical simulations
of the system of variational equations for the pulse parameters and the full complex GL
equation are given in Sec.~\ref{sec:num}. A possibility of stabilization of ES solutions
is investigated also in this section.

\section{The averaged complex GL equation}
\label{sec:aver}

    We consider the complex Ginzburg-Landau equation with nonlinearity management

\begin{eqnarray}
\label{eq:cq} i u_t + \frac{1}{2}u_{xx} + \gamma(t) |u|^2 u + \chi |u|^4 u =
i\delta u + i\beta u_{xx}
\nonumber\\
+\ i\alpha|u|^2 u + i\kappa |u|^4 u \, ,
\end{eqnarray}

where
$\gamma(t) = \gamma_0 + \gamma_1(t/\epsilon)/\epsilon$, and $\epsilon \ll 1.$ This
equation models mode-locked lasers~\cite{Ippen}, long optical communication
lines~\cite{Bak}, and a  soliton fiber lasers with two segments of alternating signs of
nonlinearity~\cite{Ilday}.  In the case of passive mode-looked fiber lasers, $u$ is the
normalized envelope of the field, $t$ is the propagation distance, $x$ is the number of
trips (retarded time), $\gamma$ and $\chi$ correspond to the Kerr and saturating
nonlinearities, $\delta $ is the linear gain (loss) coefficient, $\beta$ is the spectral
filtering, and $\alpha $ and $\kappa$ describe nonlinear gain and absorption,
respectively. We mention that a case of the rapidly varying dissipation in the framework
of the nonlinear Schrodinger equation ($\chi = \alpha = \beta =\kappa= 0$) is considered
in Ref.~\cite{Beheshti}.

   We address here the case of periodic management $\gamma_1(\tau+1)=\gamma_1(\tau),
\tau = t/\epsilon$ and $\int_0^1 \gamma_1 (\tau) d \tau =0$. We denote by $\Gamma_1$ the
zero-mean antiderivative of $\gamma_1$ i.e.
$$
\Gamma_1 = \int_0^{\tau}\gamma_{1}(\tau)d\tau - \int_0^1d\tau
\int_0^{\tau}\gamma_{1}(\tau^{\prime})d\tau^{\prime}.
$$
Following the procedure developed  in Refs.~\cite{Pelinovsky,Abdullaev03}, we average the
complex Ginzburg-Landau equation over fast variations. Following these works, we
represent field $u$ as the following
\begin{equation}
  u = ve^{i\Gamma_{1}|v|^2}.
\end{equation}
Substituting this transformation into Eq.~(\ref{eq:cq}) we obtain the following equation
\begin{eqnarray}
  iv_t &-&\Gamma_1 v(|v|^2)_t + \frac{1}{2}v_{xx} + \gamma_0 |v|^2 v
\nonumber\\
  &+& \chi |v|^4 v + 2i\Gamma_1 (|v|^2 )_x v_x +
i\Gamma_1 (|v|^2 )_{xx} v - \Gamma_1^2 (|v|^2 )_x v
\nonumber\\
  &=&i\delta v +  i\beta [v_{xx} +2i\Gamma_1 (|v|^2 )_x v_x + i\Gamma_1 (|v|^2 )_{xx}v
\nonumber\\
  &-&\Gamma_1^2 \left[(|v|^2 )_x \right]^2 v] + i\alpha |v|^2 v + i\kappa |v|^4 v.
\end{eqnarray}
We consider that the solution depends on the fast time $\tau$ and the usual $T=t$.
   Looking for the solution
$$
v(x,t) = w(x,T) + \epsilon v_{1}(x,\tau,T) + \epsilon^2 v_{2}(x,\tau,T) + \dots
$$
with $\epsilon \ll \beta, \alpha, \kappa$ and averaging over rapid oscillations we obtain
the averaged complex Ginzburg-Landau equation for $w(x,t)$ (below we use notation $T \to
t$)

\begin{eqnarray}
iw_t &+& \frac{1}{2}w_{xx} + \gamma_{0}|w|^2 w + \chi |w|^4 w + \sigma^2 \bigg[
(|w|^2)_{xx}|w|^2 \bigg.
\nonumber\\
  &+&\frac{1}{2}(|w|_x^2 )^2 \bigg. \bigg] w = R \equiv i\delta w + i\beta w_{xx} +
i\alpha |w|^2 w
\nonumber\\
&+& i\kappa |w|^4 w + i\beta\sigma^2 (|w|^2)_x w \bigg[2\left(w_x w^{\ast} -w
w_x^{\ast}\right)\bigg.
\nonumber\\
&-&(|w|^2)_x \bigg. \bigg]. \label{cgl_w}
\end{eqnarray}

Here
$$\int_0^1 d\tau \Gamma_1(\tau)=0,  \int_0^1 d\tau
\Gamma_1^2(\tau) = \sigma^2 .
$$
Note that $\sigma^2 \sim O(1)$, so we keep terms $\sim \sigma^2\beta$.
In the case of weak management with $\sigma^2 \ll 1$, these terms should be dropped~\cite{Abdullaev03}.

   Dissipative solitons exist when two balances are satisfied:
the first one is the balance between nonlinearity and dispersion, and the second balance
is between gain and loss. Comparing with the standard complex GL equation
(Eq.~\ref{eq:cq} with $\gamma (t) = const$), we observe that two effects due to the
nonlinearity management appear: the fourth order nonlinearity terms leading to the
reduction of the nonlinearity~\cite{Abdullaev09}, and the nonlinear renormalization of
the filtering term $i\beta u_{xx}$. Thus, we can expect changes in the stability regions
of dissipative solitons.

\section{ The system of equations for  parameters of a nonlinearity
managed dissipative soliton} \label{sec:breath}

   For the conservative system ($\delta = \alpha = \beta = \kappa =0$),
we have the Lagrangian density for Eq.~(\ref{cgl_w})

\begin{equation}
L = \frac{i}{2} \bigg( w_tw^{\ast}-w^{\ast}w^{\ast}_t \bigg)-\frac{1}{2}|w_x|^2
+\frac{\gamma_0}{2}|w|^4 +\frac{\chi}{3}|w|^6 -\frac{1}{2} \sigma^2 |w|^2 \left[(|w|^2)_x
\right]^2 dx. \label{lagrang}
\end{equation}
In order to derive the system of equations for the dissipative soliton parameters, we
use the variational approach~\cite{Anderson}. According to the variational approach, we should
calculate the averaged Lagrangian
$$
\bar{L}=\int_{-\infty}^{\infty} L dx,
$$
for a given ansatz for the wave profile $w(x,t)$.  We will use the Gaussian ansatz for
the dissipative soliton (DS) solution

\begin{eqnarray}
  w(x,t) = A(t)\exp \left\{-\frac{[x-x_{0}(t)]^2}{2 a^2(t)}
\right.
\nonumber\\
  \Bigg. + ik(t)[x-x_{0}(t)] + ib(t)[x - x_{0}(t)]^2 +i\phi(t) \Bigg\},
\end{eqnarray}

where $A(t),\ a(t)$ and $x_{0}$ are the amplitude, width and position of the pulse
maximum, respectively, $k(t)$ is linear phase coefficient, $b(t)$ is the chirp parameter,
and $\phi(t)$ is the phase shift.  The Euler-Lagrange equations

\begin{eqnarray}
\frac{\partial\bar{L}}{\partial\eta_i}
-\frac{d}{dt}\frac{\partial\bar{L}}{\partial\eta_{i,t}}=
\int_{-\infty}^{\infty}\left(\frac{\partial u^{\ast}}{\partial\eta_i}R
+ c.c. \right).
\end{eqnarray}

derived from the averaged Lagrangian, give a system of ordinary differential equations
(ODEs) for the dissipative soliton parameters.

   A calculation of the averaged Lagrangian $\bar{L}$ gives:

\begin{eqnarray}
\frac{\bar{L}}{N} = -\frac{a^2 b_t}{2} -\phi_t -\frac{1}{4a^2} - a^2 b^2 + \frac{\gamma_0
N}{2\sqrt{2\pi}a} \nonumber\\ + \frac{\chi N^2}{3\sqrt{3}\pi a^2} -\frac{\sigma^2
N^2}{3\sqrt{3}\pi a^4} + kx_{0,t} - k^2 .
\end{eqnarray}

Here, instead of amplitude A, we introduced the norm $N =\int_{-\infty}^{\infty}|w|^{2}dx =
\sqrt{\pi}A^2 a$. Then, from the Euler-Lagrange equations, we get

\begin{eqnarray}
N_{t} &=& 2N \left[ \delta - \beta \left(\frac{1}{2a^2} + 2 b^2 a^2 \right) -\frac{2\beta
  \sigma^2 N^2}{3\sqrt{3}\pi a^4} + \frac{\alpha
  N}{\sqrt{2\pi}a} \right.
\nonumber\\
  &+& \left. \frac{\kappa N^2}{\sqrt{3}\pi a^2} \right],
\label{ODE1} \\
a_t &=& 4ab - \frac{\alpha N}{2\sqrt{2\pi}} - \frac{2\kappa
  N^2}{3\sqrt{3}\pi a}
  + \beta \left( \frac{1}{a}-4 b^2 a^3 \right),
\label{ODE2} \\
b_t &=& \left( 1 -\frac{2\chi N^2}{3\sqrt{3}\pi } \right) \frac{1}{a^4} -4b^2
  -\frac{\gamma_0 N}{2\sqrt{2\pi} a^3}
\nonumber\\
  &+& \frac{8\sigma^2 N^2}{3\sqrt{3}\pi a^6} - \frac{4\beta b}{a^2},
\label{ODE3} \\
k_t &=& -2\beta \left( \frac{1}{a^2} + 4a^2b^2 +\frac{4N^2\sigma^2}{3\sqrt{3}a^4\pi} \right) k,
\label{ODE4}\\
x_{0,t}&=& \left(\frac{1}{2} - 4a^2b\beta \right)k,
\label{ODE5}\\
\phi_t &=& -\frac{1}{4a^2}+2b\beta + \left(\frac{1}{4} - 4a^2b\beta \right)k^2 + \frac{5N\gamma_0}{4a\sqrt{2\pi}} \nonumber\\
  &+& \frac{2N^2}{3\sqrt{3}a^2\pi} \left(2\chi - \frac{5\sigma^2}{a^2}-4b\beta\sigma^2 \right).
\label{ODE6}
\end{eqnarray}

   Firstly, we note that since $\beta > 0$, $k$  tends to zero at $t \rightarrow \infty$, so
that the velocity is zero, and $x_{0} \rightarrow \mathrm{const}$. The initial phase does
not enter other equations. Then, we can restrict ourselves by three equations for the norm
$N$, width $a$ and chirp $b$ (Eqs.~(\ref{ODE1}) - (\ref{ODE3})).
We mention the moments method~\cite{Vlasov,FAGT,Tsoy}
gives a similar system of equations.

   Fixed points (FPs) can be found by equating the right hand sides of these equations to
zero. Due to the complexity of the system it is difficult to find FPs analytically, so
FPs are found by numerically simulations. We can study the stability of these FPs by the
linear stability analysis. This approach requires the investigation of eigenvalues of the
corresponding Jacobian matrix. In numerical below we use  simulations a set of parameters
from articles~\cite{AST,Tsoy}.

\section{Numerical simulations}
\label{sec:num}

   In this section we present results of numerical simulations of the full complex GL
equation~(\ref{eq:cq}) with NM as well as of the dynamical system Eqs.~(\ref{ODE1})-(\ref{ODE3}),
 and we compare them with predictions of the variational approach. We
consider modulation of nonlinearity of the form $\gamma = 1 + f \cos(\omega t)$. Then for
$\sigma^2$ we get the expression $\sigma^2 = f^2/(2\omega^2)$

It should be noted that this system of ODEs~(\ref{ODE1})-(\ref{ODE6}) is obtained from
the averaged complex GL equation~(\ref{cgl_w}). Numerical simulations show that the
system~(\ref{ODE1})-(\ref{ODE6}) describes the dynamics of solutions of the complex
GL equation~(\ref{eq:cq}) well.

\begin{figure}[htbp]
\centerline{
\includegraphics[width=6.5cm,angle=0,clip]{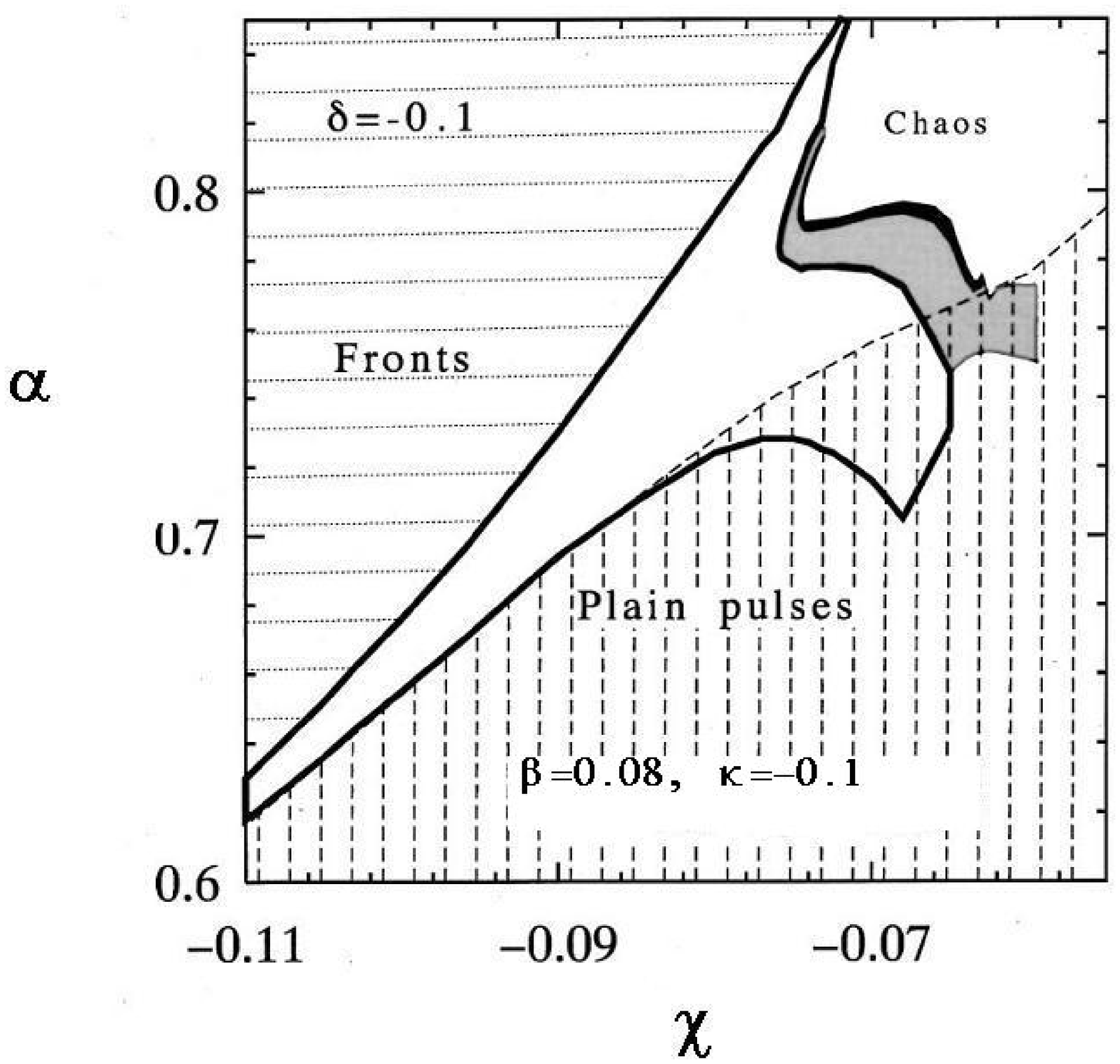}
\includegraphics[width=6.5cm,height=6.4cm,angle=0,clip]{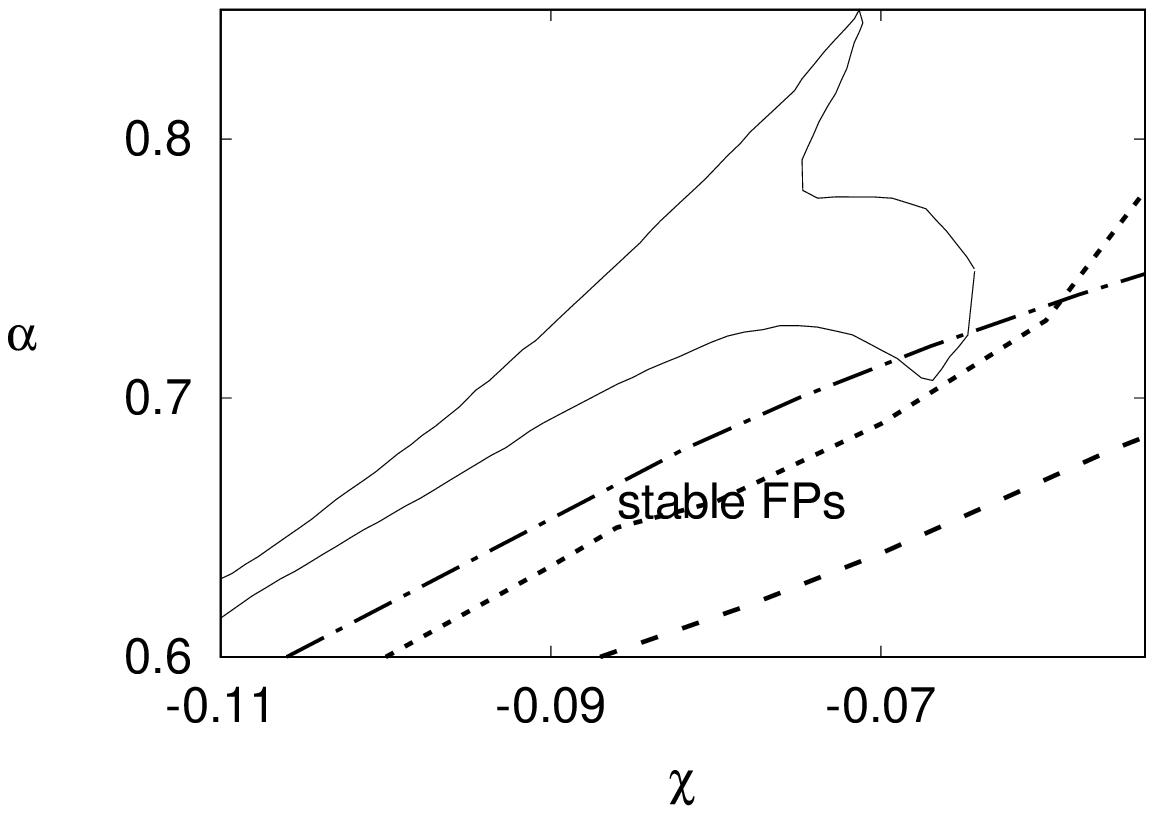}
}
\caption{(a) Regions of existence of various solutions obtained from numerical
simulations~\cite{AST} of the complex GL equation (1) without NM. The parameters are
$\delta = -0.1, \ \kappa=-0.1$  and  $\beta = 0.08,$ (b) Regions of existence and stability
of FPs with NM. The parameters of NM are: $f = 10$  and $\omega =63$.}
\label{fig:diagrm}
\end{figure}

   In the Fig.~\ref{fig:diagrm}a, regions of the existences of the different types of
solutions of Eq.~(\ref{eq:cq}) and the corresponding boundaries are presented.
Fig.~\ref{fig:diagrm}a, taken from work~\cite{AST}, shows in plane $(\chi, \alpha)$
results of numerical simulations of the original complex GL equation~(\ref{eq:cq}) for
values of parameters $\delta = -0.1$, $\kappa = -0.1$, $\beta = 0.08$ and $f = 0$. The
region with vertical shading corresponds to stationary solitons. Solutions describing
fronts exist in the region with horizontal shading. The central region between these two
regions corresponds to the pulsating solitons with the single-period variation. A small
grey area corresponds to multi-period pulsating solitons. An white area in the upper right
corner denotes region of chaotic solitons.
Stationary and pulsating solitons of Eq.~(\ref{eq:cq}) with $f = 0$ correspond to
stable FPs and LCs in dynamical model ((\ref{ODE1})-(\ref{ODE3}) with $f = 0$)(see e.g.~\cite{Tsoy}).

Stable FPs of Eqs~(\ref{ODE1})-(\ref{ODE3}) correspond to solitons with oscillating parameters are
in Eq~(\ref{eq:cq}). Oscillations of the parameters are due to NM. Stable LCs of Eqs~(\ref{ODE1})-(\ref{ODE3})
might also manifest themselves as oscillating solitons in Eq~(\ref{eq:cq}). We do not distinguish between
these two types, and all solitons with almost periodic variations of the parameters are referred as
{\em managed oscillating (MO) solitons}.

\begin{figure}[htbp]
\centerline{
\includegraphics[width=6.5cm,angle=0,clip]{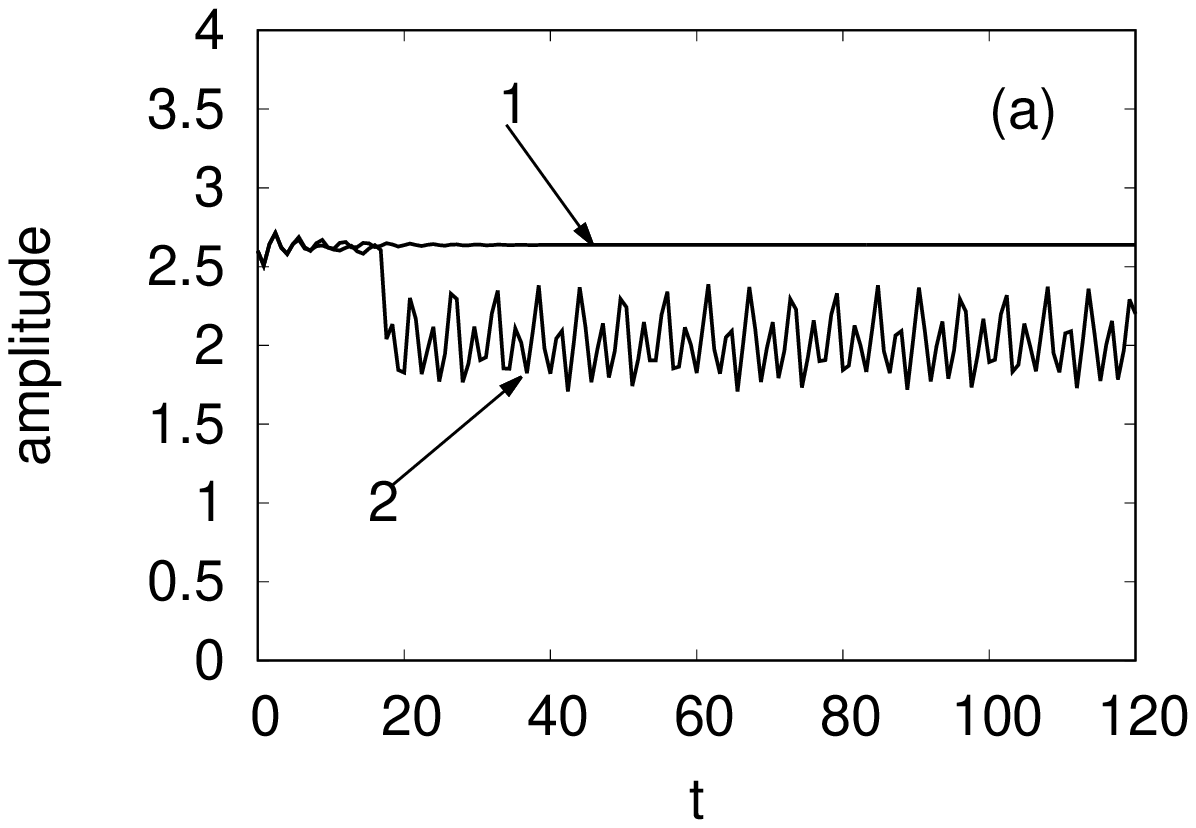}
\includegraphics[width=7.5cm,angle=0,clip]{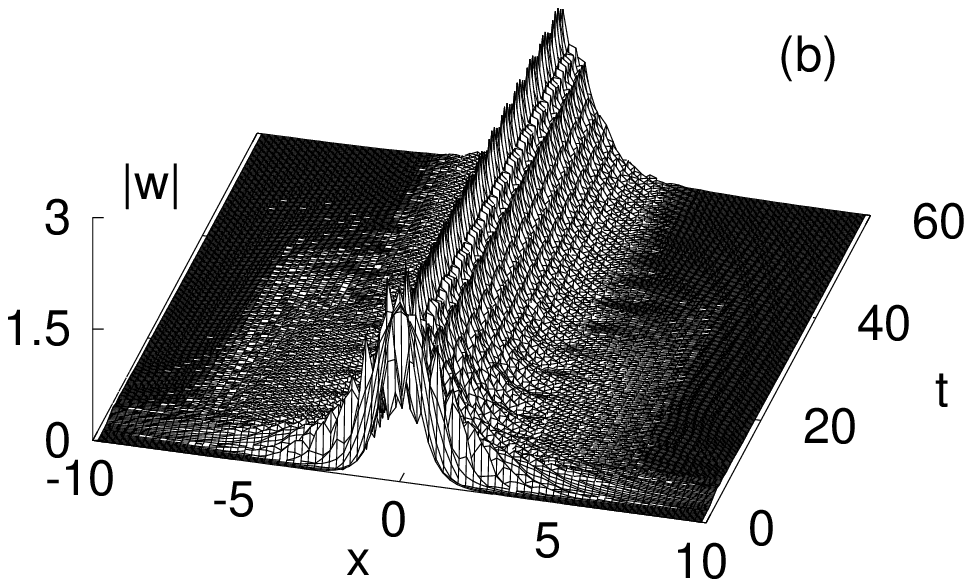}
}
\caption{(a) The pulse amplitude evolution without (line 1) and with (line 2) NM. \
(b) Evolution of the MO soliton.
 The parameters are  $\delta = -0.1, \ \kappa=-0.1, \ \beta = 0.08, \
 \chi = -0.08 \ \alpha = 0.61, \ f =10$ and  $\omega = 65$.}
\label{fig:ampl}
\end{figure}

 Figure~\ref{fig:ampl}a shows the evolution of the soliton amplitude without and with NM. We see that a
stationary soliton is transforming to a MO soliton under the action of NM. Figure~\ref{fig:ampl}b
shows the corresponding evolution of the field.

   We obtain an analogue of Fig.~\ref{fig:diagrm}a for the case, when NM is
applied, with the NM strength is $\sigma^2 = 0.013$ (or $f = 10$ and $\omega =63$).
The results obtained are displayed in Fig.~\ref{fig:diagrm}b, where for comparison
with Fig.~\ref{fig:diagrm}a, the central region is copied.

The dash-dotted line and the dashed line are obtained from Eqs.~(\ref{ODE1})-(\ref{ODE3}).
The dash-dotted line is the area boundary, below which FPs are stable in absence of NM.
Under the action of NM, this boundary is shifted down into a position which is marked by
the dashed line. The dotted line, obtained by numerical
simulations of the complex GL equation~(\ref{eq:cq}) with NM, represents the upper boundary
of MO solitons. It is seen, that the agreement between the VA system~(\ref{ODE1})~-~(\ref{ODE3})
and Eq.~(\ref{eq:cq}) is good. The region above the dotted line corresponds to
irregular solutions (fronts, chaotic solutions and others).

\begin{figure}[htbp]   
\centerline{
\includegraphics[width=6.5cm,angle=0,clip]{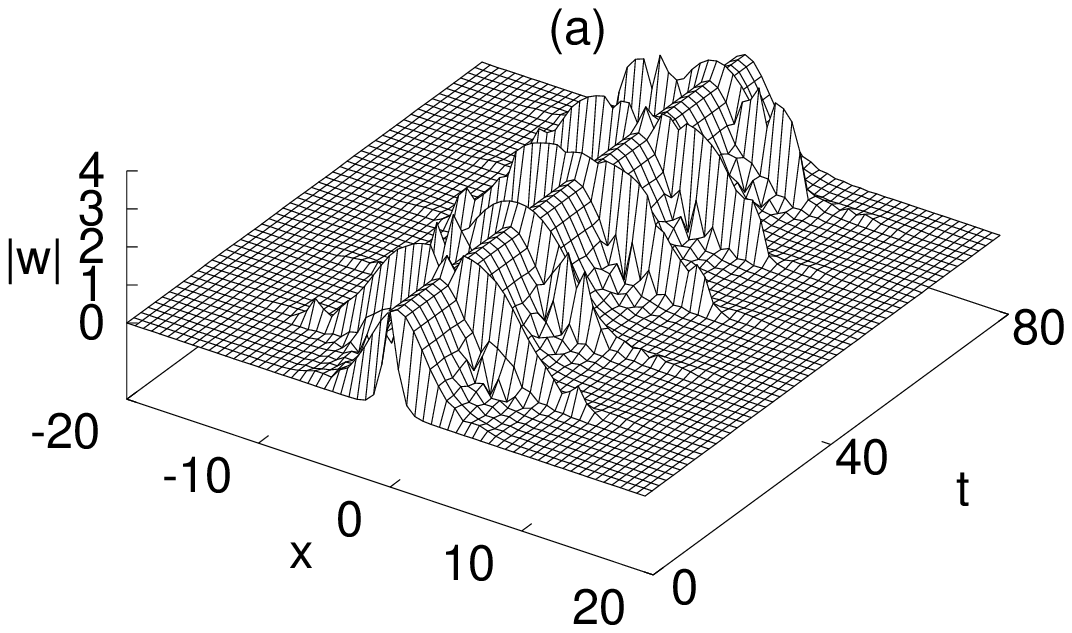}
\includegraphics[width=6.5cm,angle=0,clip]{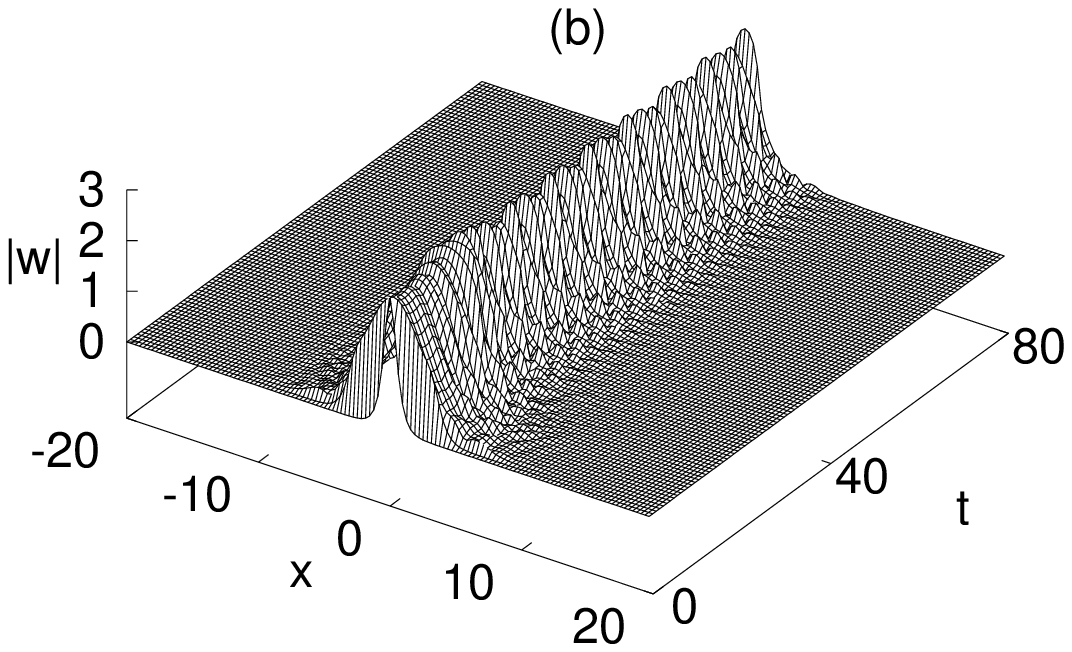}
}
\caption{(a) An exploding soliton without NM. The parameters
are $\delta = -0.1, \ \kappa=-0.1, \ \beta = 0.125, \ \chi = -0.6$ , \ $\alpha =
1.0$.
(b) An oscillating solution for the same values parameters and with NM
($f=10$ \ and \ $\omega =10$).}
\label{fig:explsr}
\end{figure}

   Numerical simulations of Eq~(\ref{eq:cq}) show that the strong management regime
($\sigma^2 \sim O(1)$) is effective for stabilization of exploding solitons solutions,
transforming them into oscillating solitons.  In Fig.~\ref{fig:explsr}(a) the
evolution of an ES in the absence of NM is presented. The evolution of solutions under NM at
the same values of parameters and for $f = 10$ and $\omega = 10$ is displayed in
Fig.~\ref{fig:explsr}(b). It is seen that the ES is transformed into a oscillating solution.

\begin{figure}[htbp]
\centerline{
\includegraphics[width=6.5cm, angle=0,clip]{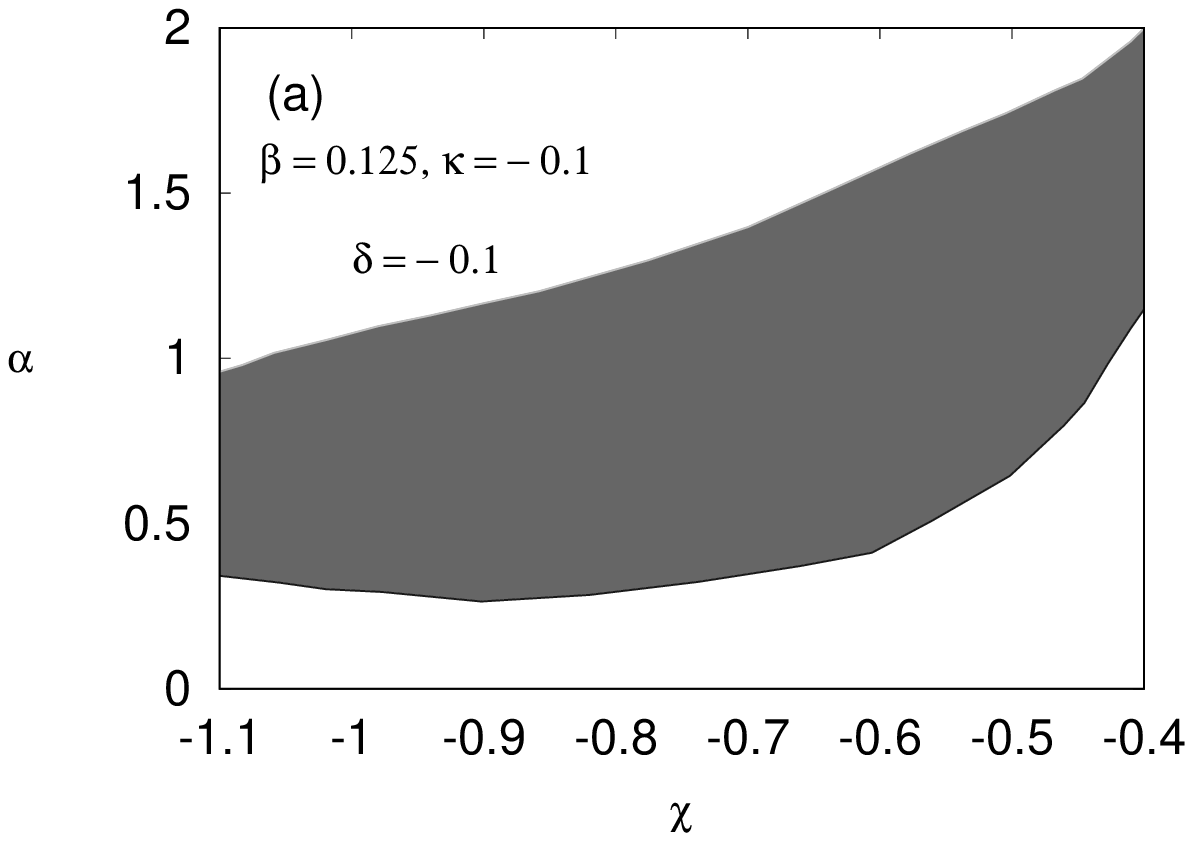}
\includegraphics[width=6.5cm,angle=0,clip]{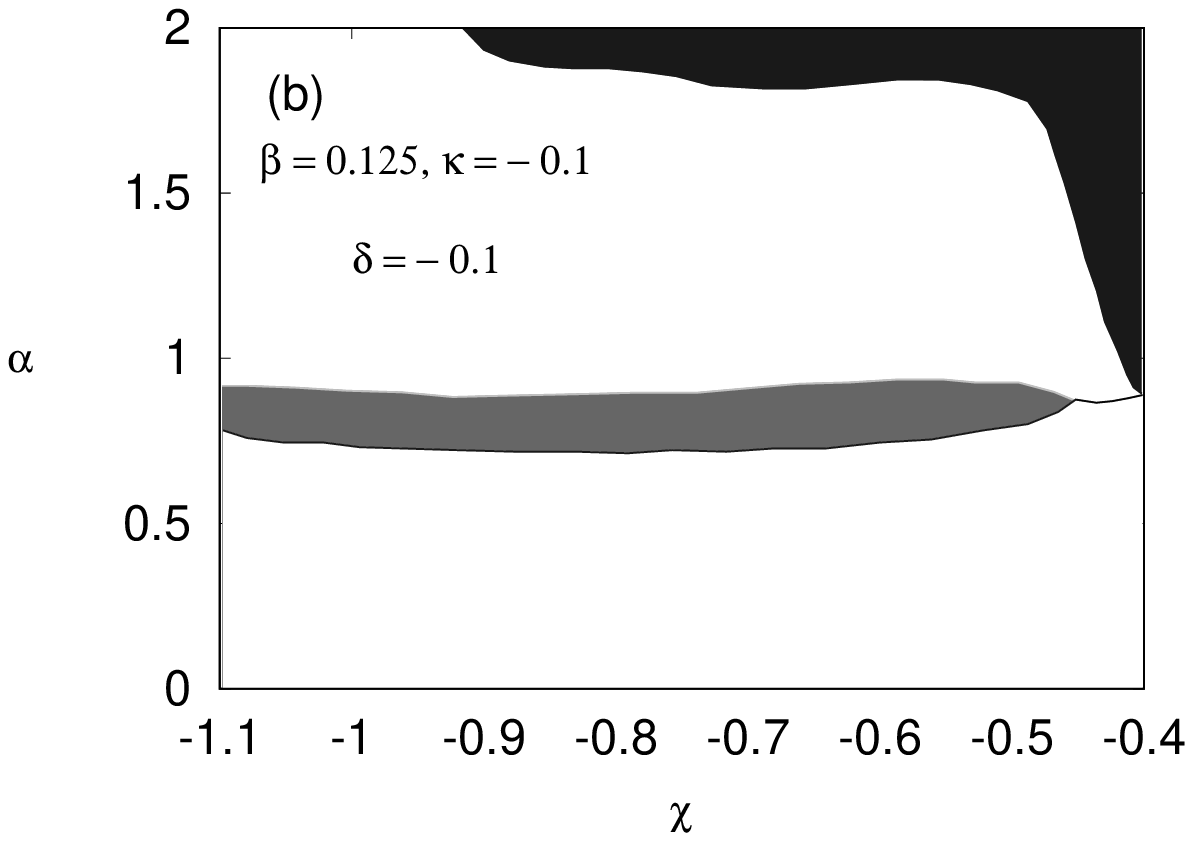}
} \caption{(a) A region in $(\alpha, \chi)$ plane different types of the the dynamics, found
exist~\cite{AST}, in the absence NM. The middle grey area corresponds to ESs. (b) Same
region when there is of NM with $f=10$ \ and \ $\omega =10$.}
\label{fig:expld_region}
\end{figure}

   In Fig.~\ref{fig:expld_region}(a), taken from work~\cite{AST}, the region of existence
   of ESs in the $(\chi, \alpha$) plane is shown. The upper white region corresponds to
   stationary solitons, while the lower white region corresponds to decaying solutions.
   The middle grey region is a region of ES solutions. We find by direct numerical
   simulations of Eq.~(\ref{eq:cq}) for $f = 10$ and $\omega = 10$ (see
   Fig.~\ref{fig:expld_region}b) that under NM these regions are deformed. As can be
   seen, the region of the ES existence is strongly reduced. Nonlinearity management
   leads to the suppression of ESs, transforming them into oscillating solutions (i.e.
   enlarging the upper white area), or into decaying solutions (enlarging lower white
   area). Also, under the action of NM a part of stationary solutions transforms into
   irregular solutions (fronts, chaotic solutions and others), see the black area in
   Fig.~\ref{fig:expld_region}b). Thus we conclude that ESs are effectively suppressed
   under the action of NM.

\section{Conclusion.}

   In conclusion, we have investigated the complex GL equation with the
strong nonlinearity management. The averaged complex GL equation has been derived.
We show the existence of nonlinearity managed dissipative solitons.
We employ the variational approach to this averaged equation and derive the system
of equations for dissipative soliton parameters.
Numerical simulations show that the explosion regime in the evolution of the
dissipative soliton is suppressed by this scheme. The region of parameters, where
explosions are suppressed, has been obtained. This type of suppression
can be observed in experiments that involve fiber-ring lasers with the
nonlinearity management.

\section{Acknowledgments}
We thank Dr. E. N. Tsoy for fruitful discussions.
The work was supposed by the grant FA-F2-004 of Ministry of Innovative Development of the
Republic of Uzbekistan.

\end{document}